# RIDGE: Reproducibility, Integrity, Dependability, Generalizability, and Efficiency Assessment of Medical Image Segmentation Models

Farhad Maleki[1,2,3*], Linda Moy[4], Reza Forghani[3], Tapotosh Ghosh[1], Katie Ovens[1], Steve Langer[5], Pouria Rouzrokh[5], Bardia Khosravi[5], Ali Ganjizadeh[5], Daniel Warren[6], Roxana Daneshjou[7,12], Mana Moassefi[5], Atlas Haddadi Avval[8], Susan Sotardi[9], Neil Tenenholtz[10], Felipe Kitamura[11], Timothy Kline[5]

[1]Department of Computer Science, University of Calgary, Calgary, AB, Canada.

[2]Department of Diagnostic Radiology, McGill University, Montreal, QC, Canada.

[3]Department of Radiology, Division of Medical Physics, University of Florida, Gainesville, FL, USA.

[4]Department of Radiology, New York University Langone Health, New York, NY, USA.

[5]Department of Radiology, Mayo Clinic, Rochester, MN, USA.

[6]Carle College of Medicine University of Illinois Urbana-Champaign, Urbana, IL, USA, Urbana, IL, USA.

[7]Department of Dermatology, Stanford School of Medicine, Stanford, CA, USA.

[8]Mashhad University of Medical Sciences, Mashhad, RK, Iran.

[9]Department of Radiology, Children's Hospital of Philadelphia, Philadelphia, PA, USA.
[10]Microsoft Research, Stanford, CA, USA.

[11]DasaInova, Diagn´osticos da Am´erica S.A, S˜ao Paulo, Brazil.

[12]Department of Biomedical Data Science, Stanford School of Medicine, Stanford, CA, USA.

*Corresponding author(s). E-mail(s): farhad.maleki1@ucalgary.ca

**Abstract**

Deep learning techniques hold immense promise for advancing medical image analysis, particularly in tasks like image segmentation, where precise annotation of regions or volumes of interest within medical images is crucial but manually laborious and prone to interobserver and intraobserver biases. As such, deep learning approaches could provide automated solutions for such applications. However, the potential of these techniques is often undermined by challenges in reproducibility and generalizability, which are key barriers to their clinical adoption. This paper introduces the RIDGE checklist, a comprehensive framework designed to assess the Reproducibility, Integrity, Dependability, Generalizability, and Efficiency of deep learning-based medical image segmentation models. The RIDGE checklist is not just a tool for evaluation but also a guideline for researchers striving to improve the quality and transparency of their work. By adhering to the principles outlined in the RIDGE checklist, researchers can ensure that their developed segmentation models are robust, scientifically valid, and applicable in a clinical setting.

## 1 Introduction

Medical images are widely used for diagnosis, monitoring, and treatment planning. To analyze these images, often a region of interest (ROI) or a volume of interest (VOI) is manually contoured by a medical expert. In diagnostic imaging, lesion segmentation enables the determination of the area or volume of a lesion of interest, including serial follow-up. It also excludes areas of normal adjacent tissue that do not provide information that is helpful with patient management. This clear delineation of the ROI/VOI is valuable both in oncology and for multiple non-oncologic applications. Segmentation is also an integral part of radiation therapy planning. Furthermore, lesion segmentation could form the first step of more advanced lesion classification tasks using traditional radiomics or deep learning approaches. However, this process is tedious, time-consuming, and prone to interobserver and intraobserver variability [1–4]. Considering the rare and highly in-demand expertise, semi-automated or automated approaches can accelerate manual segmentation and free expert time. Therefore, segmentation algorithms with the potential to be deployed in clinical settings are of high value. Automated or semi-automated segmentation approaches that are seamlessly integrated into clinical workflow can accelerate research but also can be of value in clinical practice, enabling more routine incorporation of volumetrics in clinical practice, in addition to providing the first step in more complex AI-assisted classification tasks.

Despite the many segmentation models proposed in the literature, relatively few have been deployed for clinical applications. The lack of generalizability and reproducibility of the published methodologies has been one significant obstacle to the widespread adoption of these technologies in clinical settings [5]. Many studies have highlighted that methodological deficiencies are common issues preventing the reproducibility and generalizability of prediction studies [6–8].

Several guidelines have been proposed to improve the reproducibility of predictive models in biomedical science. TRIPOD was proposed as a guideline for facilitating transparent reports of studies using multivariate predictive models for diagnosis and prognosis [9]. CONSORT was also proposed as a set of guidelines for reporting randomized controlled trial results [10]. Inspired by CONSORT, the STARD guideline was developed to enhance the reporting of diagnostic accuracy studies [11]. CLAIM was adapted from STARD to extend it to AI applications for medical imaging, including image classification, image reconstruction, and workflow optimization [12]. Although these guidelines contribute to improving the quality of reports on medical image analysis, they are not tailored for segmentation studies. Further, they mainly focus on reproducibility and provide little to no insight into developing generalizable models. As such, they lack the specific guidelines to ensure the generalizability of segmentation approaches.

Inspired by the aforementioned guidelines, this paper provides a checklist to increase the reproducibility and generalizability of machine learning-based segmentation models. The checklist aims to help authors develop more generalizable and reproducible segmentation models. It also assists reviewers in better evaluating academic manuscripts on medical image segmentation.

## 2 RIDGE Checklist

### 2.1 The Introduction Section

#### I-1: Background, purpose, and how the segmentation model will be integrated into clinical workflow

The introduction section should include the required clinical and scientific background for understanding the study and its potential applications and impact. The authors should state the clinical question they wish to address and the current standard of care for the ROI/VOI segmentation. If applicable, state-of-the-art approaches should be mentioned, as well as their drawbacks. Furthermore, the intended use of developed models or methodologies and how they contribute to the clinical workflow should be explained. Lastly, although not an absolute requirement, a paper discussing regulatory considerations (e.g. FDA for the United States), and integration into clinical workflow, along with the automated reporting of key results, would enhance the value of manuscripts, when applicable.

#### I-2: Study objectives regarding state-of-the-art segmentation models

Often, the study objective is to address some shortcomings of the state-of-the-art segmentation approaches. For example, a study objective could be to develop a segmentation model that significantly outperforms the performance of the state-of-the-art models so that the resulting model can add value in a clinical setting.

Explicitly stating a study objective(s) enables readers to better understand a study's contribution. This also sets the expectation for reviewers and facilitates manuscript assessment.

### 2.2 The Materials and Methods

*M-1: Prospective or retrospective study*

It should be indicated whether a study has been conducted prospectively or respectively. Also, some segmentation studies have both retrospective and prospective components. For example, a segmentation model can be developed and evaluated retrospectively and then further evaluated prospectively. Such cases should be appropriately documented.

*M-2: Objectives for segmentation models: development, exploration, feasibility, or comparison studies*

Whether the focus is on model creation, conducting an exploratory study, assessing feasibility, or a noninferiority trial, clear definitions and distinctions are essential. Setting a transparent goal helps readers and reviewers evaluate the technical and clinical contributions of the study and sets the context for research methodologies, results interpretation, and understanding of potential implications.

*M-3: Data sources, including imaging modality, treatment received, and protocol for image acquisition*

Comprehensive details about the imaging modality (e.g., MRI, CT) and the corresponding imaging protocol are essential for the reproducibility of a study. It is important to specify if multiple scanners and different acquisition protocols were used. It is also crucial to specify if the experiments utilized pre-treatment images, post-treatment images, or both and to articulate the rationale behind such choices. This information facilitates readers' and reviewers' understanding of the utility and clinical relevance of the proposed approach and ensures a fair comparison with state-of-the-art methods.

*M-4: Detailed information regarding the sample size used in the study*

The sample size should be explicitly mentioned. Additionally, the composition of samples across known subpopulations or subcategories of interest should be detailed. It is essential to report the sample size at the patient or participant level if multiple data points per individual are used. If relevant, the process behind its selection should be mentioned. Additionally, the sizes or proportions of data used for training, validation, and test sets should be reported.

*M-5: Eligibility criteria*

A detailed description of the process for selecting eligible participants should be provided. It is important to state where and when the potentially eligible patients have been identified. It is recommended to present this information through a flow diagram to enhance clarity. This diagram should illustrate the sequential application of each criterion and indicate the exact number of participants remaining after each step. Any potential biases introduced by the selection process and measures taken to address them should also be highlighted. Transparency in reporting key characteristics of the population of interest is not only important for generalizability but also can inform the target population for a given algorithm. A common example is providing information on whether a study included pediatric patients or not, which in turn would inform whether a tool should be used only in adults or whether it can also be used in the pediatric population.

*M-6: Detailed description of ground truth standards to allow replication of image annotations*

It is essential to provide a detailed reference standard that allows medical experts to annotate images without ambiguity. In situations where segmentation boundaries might be subjective and open to varied interpretations by different experts, a comprehensive description should be detailed to minimize interobserver variability. When multiple experts are involved in the annotation process, the methodology employed to reconcile discrepancies and arrive at a consensus for the ground truth should be explained. Alternatively, for certain segmentation tasks, some degree of variation is unavoidable. In these cases, the ground truth segmentation should be performed or edited by subspecialty-trained/certified, experienced professionals. Although there is no absolute rule, typically, segmentation from at least three independent experts is expected, although there are studies in the literature using greater than ten expert segmentations for the same patient. It may be acceptable to use one set of segmentation per patient (either performed by one or multiple experts) for training the algorithm, but the testing or evaluation should ideally be done by comparing multiple expert segmentations as described earlier. If multiple segmentations are being used to train an algorithm, then the process (e.g., use of a vote of majority-generated ROI/VOI) should be clearly reported. For testing or evaluation of the performance of a segmentation algorithm, another approach would be to report and compare the performance of the algorithm against the range of ROIs/VOIs (and consequently variations) of multiple experts.

*M-7: Justification of reference standards for ground truth image annotations*

In case boundaries of VOIs or ROIs are ambiguous, and several choices can be made for specifying them, the rationale for the choice made in the study should be provided. It is imperative to outline any potential impact the chosen reference might have on the segmentation outcomes or the study conclusions.

*M-8: Source of ground truth image annotations; qualifications and training process for annotators to generate accurate annotations*

The authors should describe the qualifications of the annotators. Also, any training or preparation provided for the annotators before contouring the images should be described. When multiple annotators contour an image, the method for handling discrepancies between these annotations should be described. Also, it is important to state if the contours have been provided manually or in a semi-automated manner, where an algorithm is used to create rough contours that are then manually edited. Lastly, as mentioned earlier, some degree of variation is unavoidable in medical segmentation tasks, and in these cases, the use of multiple expert contours is optimal to ensure reliability and generalizability.

*M-9: Tools used for image annotation*

The information about the tool(s) used for image annotation should be provided. This includes the name and the version of the software and the underlying operating system. If the annotation software provides multiple contouring tools, it is suggested that the specific ones used for image annotation be listed. When a semiautomated approach is adopted for contouring, details of the method for automatically generating the initial contours, as well as the subsequent refining process, should be described.

*M-10: Measuring and mitigating interobserver and intraobserver variability; methods for resolving annotation discrepancies*

During the data annotation phase, inconsistencies might arise due to multiple observers interpreting samples differently (interobserver variability) or a single observer providing varying annotations for the same sample on different occasions (intraobserver variability). Authors should describe the methods used to quantify interobserver and intraobserver variabilities, possibly through metrics such as Hausdorff distance, Dice coefficient, and Jaccard Index—also known as Intersection over Union (IoU). It is essential to detail any standardized guidelines, training, or protocols provided to the annotators to minimize this variability. Furthermore, the authors should outline the steps or procedures undertaken to resolve discrepancies and ensure annotation consistency.

### 2.3 Model Description

*M-11: Detailed description of model architecture, model inputs, and model outputs*

Authors should provide comprehensive details about the model architecture to ensure the model can be reconstructed based on the provided information. The expected input(s) for the model should be explicitly outlined, including image type, size, and preprocessing steps. Similarly, the expected outputs and post-processing steps must be clearly described. If feasible, a link should be provided to a public repository where the code is available.

*M-12: Strategy for initializing model parameters*

The strategy for model parameter initialization should be described. When a transfer learning approach is employed, it is essential to specify and detail the weights and biases from previously trained models. If pre-trained parameters are used, authors need to clarify which layers remain open for retraining or weight readjustment tailored to the intended task. If the model is not based on a transfer learning architecture, the method for initializing the model's parameters should be outlined.

### 2.4 Model Training

*M-13: Model hyperparameters and the methods for choosing the model hyperparameters*

Authors should describe the hyperparameters used in model training, including but not limited to learning rate, optimizer, and loss function. If hyperparameters are determined through a trial-and-error process, this procedure should be described, illustrating the range tested and the criteria for final selection. In cases where systematic hyperparameter tuning methods like grid search, random search, or Bayesian optimization are used, details of the search strategy and results should be included.

*M-14: Image preprocessing steps*

Image processing is often important in machine learning and deep learning applications. Preprocessing steps, if any, should be described with enough detail to allow for reproducibility of the results. These could include steps such as applying intensity normalizations, image cropping, or resizing.

*M-15: Image augmentation*

Image augmentation is a common practice in developing deep learning-based segmentation models, specifically in the absence of large-scale annotated datasets [13–15]. In the context of image segmentation, data augmentation refers to methods that allow for computationally transforming an image so that the image annotation for the newly generated image can be computationally inferred, alleviating the need for further data collection or manual annotation (see Figure 1). Image augmentation alleviates overfitting by introducing data variability and artificially increasing sample size. Often, a stochastic pipeline of image augmentation is composed, where a sequence of image augmentations, each with a given probability of being applied, is used to create an augmented version of an input image and its corresponding mask.

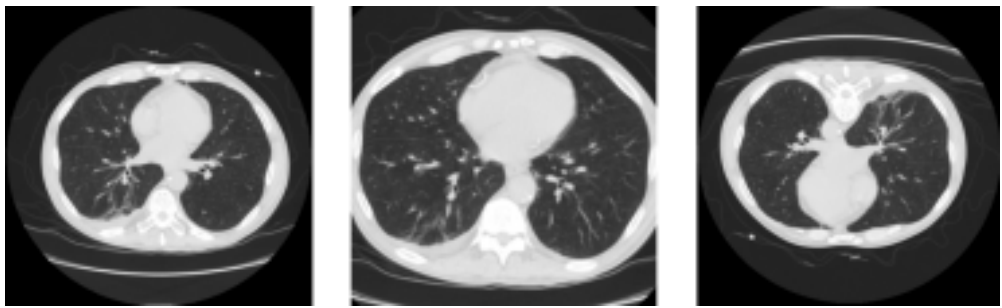

Fig. 1 Example of image augmentations. The original image (left) has been augmented by zooming (middle) and rotation (right).

To reproduce a data augmentation pipeline, it is essential to provide detailed information about the pipeline. Model-based data augmentation relies on a model to generate synthetic images [15]. When using model-based data augmentations, it is essential to provide information on how to acquire the model and instructions on how to use these models for image augmentation.

*M-16: Criteria and process for final model selection*

Model training is an iterative process where a model is updated over multiple epochs. Therefore, it is essential to establish and report criteria for selecting the best model from those developed across different epochs. This selection process can be informed by various performance metrics and stopping criteria. Consequently, reporting the specific performance metrics and the criteria used to halt the training process is crucial. A common approach, for instance, involves monitoring the loss function on the validation and training sets. Decisions can be based on a predetermined threshold for performance improvement on a validation set, a predetermined number of epochs without improvement, or other domain-specific criteria. Understanding these factors offers insights into the model's reliability and its applicability to real-world scenarios.

*M-17: Hyperparameters that led to the best model*

Unlike model parameters that are learned based on the training set during the training process, the hyperparameters are often either manually assigned or selected based on some heuristic methods. When using heuristics to choose the model hyperparameters, the validation set is used to select hyperparameters that result in better model performance. Examples of hyperparameters are learning rate, optimization algorithm, momentum, and batch size. The set of model hyperparameters that lead to the best result should be stated in the paper.

*M-18: Ensemble techniques: Model diversity, prediction consolidation, and computational considerations (if applicable)*

Due to their high capacity for learning complex problems, deep learning models often exhibit high variance, especially when trained on small datasets. Ensemble approaches combine multiple models to reduce this prediction variance and enhance overall performance [4, 5]. The individual models within an ensemble may vary in their model architectures or in the datasets on which they have been trained. When employing ensemble techniques, it is important to outline these differences. Additionally, the method by which predictions from these models are consolidated to form a final prediction should be clearly described. Also, detailed information about the added computational burden of deploying these models should be provided. The computational requirement might run an approach impractical in some clinical settings.

**2.5 Model Evaluation**

*M-19: Metrics for evaluating model performance*

Providing correct metrics for model evaluation is essential in developing generalizable models. For example, for problems where the area/volume of interest is a small fraction of the image, pixel/voxel accuracy does not provide a meaningful measure of model performance, as a model that predicts all pixels/voxel not belonging to the area/volume of interest still achieves a high accuracy value. Dice score and intersection over Union (IoU) are the most common measures used for model evaluation. To facilitate model comparison, providing these measures for all experiments is recommended. Also, for problems where other metrics, such as distance-based metrics, are commonly used to assess model performance, those should also be reported [6, 7].

*M-20: Measuring robustness or sensitivity analysis*

Robustness refers to the ability of a model to maintain consistent performance despite minor perturbations or changes in input data. Noise and artifacts are common in medical imaging; therefore, segmentation models should be robust in order to be deployed in a clinical setting. Sensitivity analysis for a given model assesses the extent to which variations in input data affect the predictions made by the model. Given the diversity of human anatomy and the variability in medical imaging modalities, understanding which factors most influence model performance can offer key insights into its potential limitations and areas for improvement. Several best practices are recommended to ensure robust performance in medical imaging models. It is important to use established quantitative metrics such as the Dice coefficient and IoU to measure how model performance changes with variation in an input image. Visual comparisons could provide a qualitative sense of model predictions across different levels of image perturbation and noise. It is essential to report both successful and unsuccessful outcomes, as understanding model limitations is especially critical in clinical contexts. Models should also be tested using images from various sources and patient groups to ascertain widespread usability. Lastly, all assumptions about input data made during analysis should be explicitly documented to highlight their potential impact on the model's clinical performance.

*M-21: Internal validation, external validation, or both*

In internal validation, a subset of the dataset is used for model training and another subset for model evaluation. In contrast, the model is evaluated using an independently derived dataset in external validation. An external dataset often provides a better estimate of model generalizability and should ideally be the primary method for model evaluation whenever possible. It should be explicitly stated whether the model evaluation is internal, external, or both.

*M-22: Level at which training, validation, and test sets are disjoint (e.g., patient or institution)*

When developing deep learning models, data is often partitioned into training, validation, and test sets. Partitions should ideally be conducted at the patient or institution level to ensure the same subject does not appear in more than one subset.

Data partitioning at the institution level can further enhance model generalizability across different setups and data sources. However, when data from different institutions systematically differ, the model performance measures on these sets might substantially vary. In such cases, these differences should be studied and reported.

*M-23: Data points for each subject are exclusively present in training, validation, or test sets*

Due to substantial anatomical similarities between different images from the same patient, a model could associate irrelevant anatomical characteristics to an endpoint of interest instead of learning the condition under study. Consequently, the performance measures do not reflect the true model performance. In the context of segmentation models, this can lead the model to memorize the segmentation map for a patient based on these unrelated characteristics. Therefore, to avoid such issues, data points from the same patient should be confined to just one of the training, validation, or test sets.

*M-24: Oversampling is not applied before splitting data into training, validation, and test sets*

Oversampling can contribute to developing segmentation models for imbalanced datasets, particularly for rare pathologies or conditions. However, if oversampling is performed before dataset partitioning, there is a risk that identical images could be distributed across the training, validation, and test sets. This would allow the model to memorize the ROIs/VOIs, resulting in a misleadingly over-optimistic assessment of the segmentation model. Therefore, it is essential to partition the dataset and then apply oversampling on the minority class(es) in the training set. This approach ensures that the model is trained on a more balanced dataset without compromising the integrity of the evaluation process.

*M-25: Image augmentation is not applied before splitting data into training, validation, and test sets*

Data augmentation should only be applied after splitting the dataset into training, validation, and test sets. Although when applying image augmentation, some of the characteristics of the augmented image change, the resulting image still retains a substantial amount of information with the original image (see Figure 1). Consequently, a model could achieve high-performance measures by memorizing segmentation maps, leading to an overestimation of performance measures and models that lack generalizability. By performing augmentation only on the training set, the overall performance of the model can be improved, without compromising the integrity of the evaluation process.

*M-26: Samples in the test set are not used to make a decision about preprocessing, model training, or post-processing*

Samples in the test set should not be used for selecting preprocessing or post-processing steps or during model training, or post-processing. Failing to adhere to this guideline can prevent the test set from providing an unbiased estimate of the model's generalization error. This oversight might lead to over-optimistic performance measures that do not accurately represent the performance of the model on unseen data.

*M-27: Describing demographic and clinical characteristics of training, validation, and test sets*

Demographic and clinical characteristics of samples in training, validation, and test sets should be described better to evaluate the clinical utility of a proposed model and to enhance the reproducibility of the results. For example, age groups (e.g., pediatric vs. geriatric populations) can have significant differences in anatomy, affecting how ROIs/VOIs appear on a medical image. Also, a trained model might perform differently for patients with different treatment histories or disease subtypes, resulting in substantially different performance measures for different compositions of test sets.

*M-28: Strategies to enhance segmentation model robustness to common image variations*

Image variations, which are inherent in medical imaging due to factors like diverse acquisition protocols, hardware differences, and software discrepancies, must be effectively managed by a model to ensure its reliability and deployability in a clinical setting.

To address these variations, techniques such as data augmentation and domain adaptation can be employed. Domain adaptation techniques can help the model generalize across different imaging settings by aligning the feature distributions from different domains, ensuring the model performs consistently well regardless of the source of the images. For example, this could enable models to perform well in the presence of image artifacts, noise, or systematic variations in image acquisitions.

*M-29: Software libraries, frameworks, and packages*

Libraries, packages, and frameworks used for training and evaluating the model(s) should be described with enough detail to allow for reproducing the results.

*M-30: Availability of trained model and the inference code for segmenting ROIs/VOIs in an image provided in a standard format, except when restricted by intellectual property considerations*

The trained model and inference code should preferably be accessible online, enabling readers and reviewers to assess the performance of the developed model with their own datasets or samples, facilitating comparison with current and future research. This accessibility should include both the preprocessing and post-processing pipelines.

## 2.6 The Results Section

*R-1: Estimates of performance measures and their variability*

We recommend providing a comprehensive report on the performance of proposed medical image segmentation model. This encompasses not just the primary metric for assessing the performance of the segmentation model, such as Dice or IOU, but also the confidence intervals that reflect the uncertainty of these measures. Furthermore, given the diversity of medical imaging conditions and modalities, it is essential to highlight potential fluctuations in model performance across known subpopulations or subcategories. Factors to consider include variations in patient demographics, imaging equipment, imaging protocols, and external interferences or noise.

*R-2: Failure analysis of poorly segmented cases*

Variations in image quality, human anatomy, and pathology, as well as overlapping structures, often lead to errors in predictions made by segmentation models. Segmentation errors typically manifest as false positive segments, false negative segments, or boundary inaccuracies. For example, when normal tissue is predicted as a tumor, it is a false positive segment; when the model partially or completely misses a tumor, the missing part is a false negative segment.

Often, a single measure is used to describe model performance. However, this approach can prevent a comprehensive understanding of model errors, especially in the presence of systematic errors. For instance, when images feature several ROIs of varying sizes, a model that misses small ROIs but accurately identifies large ones could still achieve a high aggregate score, such as IoU or Dice score. This can be misleading, as the model might be medically unreliable. Analyzing these errors offers insights that can guide model refinement. We recommend visualizing examples where a model fails to perform a medically desirable segmentation. The model errors could be assessed quantitatively or qualitatively.

*R-3: A scatter plot representing the distribution of the size of region(s) or volume(s) of interest for training, validation, and test sets*

We recommend visually assessing the distribution of sizes of the ROI(s) or VOI(s) across the training, validation, and test datasets. Each ROI or VOI can be represented as a point in this scatter plot. The Y-axis of the plot indicates the size/volume of an ROI/VOI, and different colors can be used to represent samples in training, validation, or test sets. This visualization provides insights into potential biases or imbalances in the size distribution of ROIs or VOIs. A balanced and overlapping distribution across the training, validation, and test sets suggests that the model has been trained and evaluated on a representative sample, minimizing the risk of overfitting to a specific range of ROI/VOI sizes or compromising model generalizability. Moreover, examining the overall size distribution of VOIs or ROIs can highlight the clinical utility of these models. For instance, a model primarily trained to detect large lymph nodes might have limited clinical relevance. A visual exploration of the size of ROIs/VOIs can quickly pinpoint such issues. These scatter plots can also highlight potential biases related to various confounders, such as imaging hardware, software, protocol, patient demographics, or medical conditions. This can be achieved by utilizing different shapes or colors for data points representing samples from each category of potential confounders.

Bland-Altman plots and MA plots can also be used to evaluate the discrepancies between model predictions and the ground truth regarding the size of ROIs/VOIs. A Bland-Altman plot, also known as a Tukey mean-difference plot, visualizes the difference between the two measurements against their mean. The MA plot is essentially the Bland-Altman plot for log-transformed values. MA plots use log-transformed values to depict variations from mean values.

*R-4: Analyze bias across patient categories such as relevant sociodemographics and imaging protocols and hardware*

A model might not achieve the same performance level across all patient population subgroups. These subgroups could be defined based on sociodemographic characteristics such as age or sex, or based on factors such as imaging protocol, imaging hardware, or disease type, to name a few. It is essential to assess the performance of a model across medically relevant groups to avoid bias. By offering a detailed performance breakdown by these categories, we can enhance comprehension of the strengths and weaknesses of a model. Additionally, examining images from diverse categories provides a more comprehensive view of the model's performance.

*R-5: Failure analysis by visualizing the worst-performing cases of the model in the internal test set and, if applicable, in the external test set*

A visual review of the most inaccurate predictions from the internal test set and, if applicable, the external test set is highly recommended to rapidly pinpoint areas that need refinement and assist with model improvement. It also uncovers and assists in mitigating inherent biases that the model might have. This rigorous analysis assists readers and reviewers in understanding where a model might falter and informs us about its trustworthiness. Further, by examining potential failures on external datasets, we can verify the ability of a model to generalize across diverse scenarios.

*R-6: Performance on external dataset(s), if possible, and explaining any statistically significant difference between performance measures for samples in the internal and external test sets*

Evaluating the performance of medical image segmentation algorithms on external datasets is crucial to ensure their generalizability across different data sources and conditions. Solely relying on internal datasets may lead to scenarios where a model performs exceptionally well on one specific dataset but fails in real-world scenarios. Statistically assessing any significant discrepancies between the performance of a model on the internal and external test set(s) provides insights into potential biases, limitations, or the robustness of the model, ensuring safer and more reliable clinical application.

### 2.7 The Discussion Section

#### D-1: Study limitations, potential biases, and generalization concerns

Limitations of the study should be clearly detailed in the Discussion section. Potential biases, such as over-representation or under-representation of specific conditions or demographics that might have emerged during data collection, must be emphasized. If there is a lack of external evaluation, or if the external test set may not comprehensively represent the entire population, these issues should be explained. Additionally, any limitations related to study design, data quality, limitations related to the ground truth, or model implementation must be succinctly articulated.

#### D-2: Practical utility and clinical integration of segmentation models

The authors should discuss the practical utility of their model in a medical context. This will help readers and reviewers grasp the significance of the work and understand how it can potentially be integrated into medical practice. Discussion of adequacy or potential challenges to clinical deployment from a regulatory perspective and how the algorithm might be integrated into the workflow for clinical practice implementation would be desirable for a high-impact comprehensive study.

#### D-3: Highlighting data imbalance due to differences in the size of ROIs/VOIs and its potential effect on performance measures

If there is any data imbalance resulting from varied ROI/VOI sizes, the authors should clearly highlight this. They should also articulate the impact of such imbalance on the performance of the proposed model, including potential improvements or deteriorations if the dataset were balanced. Furthermore, the authors need to discuss the measures they took to address this imbalance, thereby demonstrating the value of their work.

### 2.8 The Conclusion Section

#### C-1: Concise presentation

The authors should succinctly describe the contributions of their work in this section. This may include the novelty of the proposed approach, primary contributions, a brief overview of the methodology, the most notable findings of the study, and potential implications or future directions for research.

#### C-2: Proper positioning of this work in the context of state-of-the-art practice, if applicable

Proper positioning of research within the context of state-of-the-art practices provides readers and reviewers with a clear understanding of how the presented work compares with or advances beyond the current best practices in the field. In instances where particular research does not necessarily surpass the state-of-the-art, it is still important to understand its position in the broader landscape, as this can illuminate complementary aspects, potential synergies, or alternative perspectives.

*C-3: Recommendations for future work, if applicable*

To guide subsequent studies, it is recommended that the authors detail challenges faced, highlight unaddressed gaps, and suggest potential methods or refinements for the future. This section should also hint at broader areas warranting exploration based on current findings, propose practical real-world applications, and, if applicable, provide an overview of planned subsequent research. By doing so, the paper paves the way for future scholarly endeavors.

*C-4: The conclusion is adequately supported by the results of the study*

It is essential to ensure that the conclusions of a study are directly derived from, and supported by, the empirical findings presented in the paper. It is imperative that there be no overgeneralization of results. By ensuring that the conclusions are firmly grounded in the actual results, the study preserves its integrity, credibility, and relevance to its audience.

### 2.9 Source Code

*S-1: Code is made available, or if not, is justified within the manuscript as to why it is omitted*

Source code greatly assists readers in comprehending the methodology detailed in the articles. By executing the code in their own environments, readers can conduct experiments with new data using the existing methodology or potentially leverage it to develop new methodologies. This availability of code also facilitates comparisons with future research and the state-of-the-art in the field. Thus, it is recommended that the source code be made accessible and referenced in the article. If, for any reason, the code cannot be made available, authors should articulate the reasons for its omission.

| |
|---|
| **I-1 Background, purpose, and how the segmentation model will be integrated into clinical workflow** |
| **I-2 Study objectives regarding state-of-the-art segmentation models** |
| **M-1 Prospective or retrospective study** |
| **M-2 Objectives for segmentation models: development, exploration, feasibility, or comparison studies** |
| **M-3 Data sources, including imaging modality, treatment received, and protocol for image acquisition** |
| **M-4 Detailed information regarding the sample size used in the study** |
| **M-5 Eligibility criteria** |
| **M-6 Detailed description of ground truth standards to allow replication of image annotations** |
| **M-7 Justification of reference standards for ground truth image annotations** |
| **M-8 Source of ground truth image annotations; qualifications and training process for annotators to generate accurate annotations** |
| **M-9 Tools used for image annotation** |
| **M-10 Measuring and mitigating interobserver and intraobserver variability; methods for resolving annotation discrepancies** |
| **M-11 Detailed description of model architecture, model inputs, and model outputs** |
| **M-12 Strategy for initializing model parameters** |
| **M-13 Model hyperparameters and the methods for choosing the model hyperparameters** |
| **M-14 Image preprocessing steps** |
| **M-15 Image augmentation** |
| **M-16 Criteria and process for final model selection** |
| **M-17 Hyperparameters that led to the best model** |
| **M-18 Ensemble techniques: Model diversity, prediction consolidation, and computational considerations (if applicable)** |
| **M-19 Metrics for evaluating model performance** |
| **M-20 Measuring robustness or sensitivity analysis** |
| **M-21 Internal validation, external validation, or both** |

**M-22 Level at which training, validation, and test sets are disjoint (e.g., patient or institution)**

**M-23 Data points for each subject are exclusively present in training, validation, or test sets**

**M-24 Oversampling is not applied before splitting data into training, validation, and test sets**

**M-25 Image augmentation is not applied before splitting data into training, validation, and test sets**

**M-26 Samples in the test set are not used to make a decision about preprocessing, model training, or post-processing**

**M-27 Describing demographic and clinical characteristics of training, validation, and test sets**

**M-28 Strategies to enhance segmentation model robustness to common image variations**

**M-29 Software libraries, frameworks, and packages**

**M-30 Availability of trained model and the inference code for segmenting ROIs/VOIs in an image provided in a standard format, except when restricted by intellectual property considerations**

**R-1 Estimates of performance measures and their variability**

**R-2 Failure analysis of poorly segmented cases**

**R-3 A scatter plot representing the distribution of he size of ROIs/VOIs for the training, validation, and test sets**

**R-4 Analyze bias across patient categories such as relevant sociodemographics and imaging protocols and hardware**

**R-5 Failure analysis by visualizing the worst-performing cases of the model in the internal test set and, if applicable, in the external test set**

**R-6 Performance on external dataset(s), if possible, and explaining any statistically significant difference between performance measures for samples in the internal and external test sets**

**D-1 Study limitations, potential biases, and generalization concerns**

**D-2 Practical utility and clinical integration of segmentation models**

**D-3 Highlighting data imbalance due to the differences in the size of ROIs/VOIs and its potential effect on performance measures**

**C-1 Concise presentation**

**C-2 Proper positioning of the work in the context of state-of-the-art practice, if applicable**

**C-3 Recommendations for future work, if applicable**

**C-4 The conclusion is adequately supported by the results of the study**

**S-1 Code is made available, or if not, is justified within the manuscript as to why it is omitted**

# 3 Discussion

In this paper, we proposed the RIDGE checklist, a comprehensive framework designed to enhance the reproducibility and generalizability of biomedical image segmentation models. The RIDGE checklist aims to fill the gaps left by existing guidelines, such as CLAIM, by addressing the unique challenges associated with segmentation tasks in medical imaging. By incorporating detailed criteria and best practices, RIDGE provides researchers and reviewers with a structured approach to ensure that segmentation models are scientifically robust and clinically applicable. RIDGE covers various aspects of model development, from data handling and augmentation techniques to evaluation metrics and bias analysis, thereby facilitating the development and evaluation of reliable and generalizable segmentation models that can be effectively integrated into clinical workflows. Table 1 provides the checklist where guidelines are referred to by a character representing their associated section—Introduction (I), Methods (M), Results (R), Discussion (D), Conclusion (C), and Source Code (S)—followed by a number.

Various checklists—such as CLAIM, STARD, and TRIPOD—have been proposed to improve the reproducibility of research in the medical domain. As these checklists collect and highlight best practices and guidelines, they have substantial overlap. The CLAIM checklist is based on STARD, which in turn was developed based on the CONSORT checklist. There is substantial overlap between different checklists as they incorporate many best practices applicable across different scenarios (see Table S1 in the Supplementary Materials). However, each checklist is designed to cater to specific needs without rendering previous checklists obsolete. Instead, they include specialized guidelines to answer a particular need. RIDGE is also no exception and aims to collect a comprehensive but not cumbersome set of best practices. RIDGE is specifically designed to include and clarify items that enhance the reproducibility of biomedical image segmentation models. While CLAIM provides a robust framework for AI applications in medical imaging, RIDGE introduces additional criteria specifically tailored for segmentation models to address their unique challenges. For instance, criteria such as M-24 and M-25 highlight the importance of applying oversampling and augmentation only after data splitting to prevent data leakage and ensure valid performance metrics. M-26 underscores the necessity of keeping test set samples separate from the decision-making process in preprocessing, model training, or post-processing to maintain an unbiased evaluation of the model's generalization capability. M-28 focuses on strategies to enhance model robustness to common image variations. RIDGE also includes criteria such as R-3, which advocates for visualizing the distribution of ROI/VOI sizes to identify potential biases, and R-4, which involves analyzing biases across different patient categories and imaging conditions. Finally, D-3 highlights the importance of addressing data imbalances related to ROI/VOI sizes to ensure robust and reliable segmentation models. These additional criteria are critical for developing clinically applicable and generalizable segmentation models.

We assessed RIDGE's efficacy, usability, and overall value in providing a structured and comprehensive framework for evaluating medical image segmentation articles by evaluating a corpus of already published papers [16–34]. Evaluating the generalizability of a model demands substantial experimentation, and our evaluation has not been aimed at reproducing these works or assessing their generalizability. Furthermore, the lack of evidence on following a guideline cannot be inferred as a failure to adhere to that guideline.

Our review of these published works showed that most papers miss one or more critical criteria regarding study reproducibility and model generalizability in their Methodology section. These often omitted or inadequately addressed criteria such as the rationale for choosing the reference standard (M-7), the measurement of and mitigation strategy for interobserver and intraobserver variability (M-10), measuring robustness or sensitivity analysis (M-20), and ensuring that oversampling is not applied before splitting data into training, validation, and test sets (M-24). Additionally, the fact that samples in the test set should not be used to make decisions about preprocessing, model training, or postprocessing is often not stated directly and cannot be inferred indirectly (M-26). The criteria related to the demographic and clinical characteristics of samples in training, validation, and test set (M-27) and the strategies to enhance segmentation model robustness to common image variations (M-28) are often omitted or not adequately addressed.

Our review also identified several areas frequently overlooked or insufficiently addressed in the Result section of these studies. Notably, the distribution of size of ROIs/VOIs across training, validation, and test sets (R-3)—which could facilitate understanding model performance and potential biases—is often not qualitatively or quantitatively highlighted. Additionally, there is a conspicuous gap in the analysis of bias, particularly regarding patient demographics, imaging protocols, and hardware variations (R-4). These could lead to models that lack generalizability and are potentially biased. Moreover, there appears to be a lack of detailed failure analysis, especially in terms of visualizing the worst-performing cases within internal and, where applicable, external test sets (R-5). This type of analysis is crucial for identifying and addressing model weaknesses. Furthermore, the performance of models on external datasets, along with an explanation for any significant differences observed between internal and external performance measures (R-6), is often not reported. If provided, such information could provide a strong signal regarding the generalizability of a model and its utility in clinical settings.

These findings highlight the need for more rigorous and comprehensive reporting to enhance the reliability and applicability of medical image segmentation models in clinical settings.

RIDGE has been primarily designed and evaluated for radiological image segmentation models. While we expect that many of the concepts are generalizable to other biomedical image segmentation tasks, such as histopathological images, we suggest that the evaluation of RIDGE for these applications be considered as future research.

## 4 Conclusion

In this manuscript, we proposed the RIDGE checklist to provide a comprehensive set of criteria for evaluating medical image segmentation studies. The RIDGE checklist has the potential to significantly enhance the quality and consistency of research in

AI-based segmentation approaches. By emphasizing a thorough review of methodologies, results, and discussions, RIDGE encouraged a more rigorous and transparent approach to study design and reporting. This is particularly crucial in medical image segmentation, which directly impacts clinical outcomes, such as in radiotherapy planning or diagnostic accuracy. The RIDGE checklist can be crucial in guiding researchers toward higher standards in AI-based medical image segmentation studies.

# Supplementary Material

Table S1. The RIDGE Checklist builds upon the best practices and guidelines already highlighted by other checklists, such as CLAIM, STARD, and TRIPOD.

| RIDGE | CLAIM | STARD | TRIPOD | RIDGE | CLAIM | STARD | TRIPOD |
|-------|-------|-------|--------|-------|-------|-------|--------|
| I-1  | ✓ | ✓ | ✓ | M-22 | ✓ |   |   |
| I-2  | ✓ | ✓ | ✓ | M-23 | ✓ |   |   |
| M-1  | ✓ | ✓ | ✓ | M-24 |   |   |   |
| M-2  | ✓ | ✓ | ✓ | M-25 |   |   |   |
| M-3  | ✓ | ✓ | ✓ | M-26 |   |   |   |
| M-4  | ✓ | ✓ | ✓ | M-27 | ✓ | ✓ |   |
| M-5  | ✓ | ✓ | ✓ | M-28 |   |   |   |
| M-6  | ✓ | ✓ |   | M-29 | ✓ |   |   |
| M-7  | ✓ | ✓ |   | M-30 |   |   |   |
| M-8  | ✓ |   |   | R-1  | ✓ |   |   |
| M-9  | ✓ |   |   | R-2  |   |   |   |
| M-10 | ✓ |   | ✓ | R-3  |   |   |   |
| M-11 | ✓ | ✓ | ✓ | R-4  |   |   |   |
| M-12 | ✓ |   |   | R-5  |   |   |   |
| M-13 | ✓ |   |   | R-6  |   |   |   |
| M-14 | ✓ |   |   | D-1  | ✓ | ✓ | ✓ |
| M-15 | ✓ |   |   | D-2  | ✓ | ✓ |   |
| M-16 | ✓ |   |   | D-3  |   |   |   |
| M-17 | ✓ |   |   | C-1  |   |   |   |
| M-18 | ✓ |   |   | C-2  |   |   |   |
| M-19 | ✓ | ✓ | ✓ | C-3  |   |   |   |
| M-20 | ✓ | ✓ |   | C-4  |   |   |   |
| M-21 | ✓ |   |   | S-1  |   |   |   |